\pdfoutput=1
\documentclass[preprintnumbers,amsmath,amssymb,prd,notitlepage,nofootinbib,twocolumn,superscriptaddress,table,xcdraw,floatfix]{revtex4-2}
\usepackage{graphicx}
\usepackage{dcolumn}
\usepackage{bm}
\usepackage{mathtools}
\usepackage{wasysym}
\usepackage{adjustbox}
\usepackage{ dsfont }
\usepackage{verbatim}
\usepackage[dvipsnames]{xcolor}
\usepackage{hyperref}
\usepackage{pbox}

\hypersetup{
	colorlinks=true,
	linkcolor=red,
	citecolor=blue,
}

\usepackage{float}
\usepackage{xspace}
\usepackage{tabularx}
\usepackage[font=footnotesize, caption=false]{subfig}
\renewcommand{\selectlanguage}[1]{}
\usepackage{blindtext}
\usepackage{verbatim}
\usepackage{mathrsfs}

\usepackage{tipa} 

\usepackage{colortbl}

\mathchardef\mhyphen="2D

\makeatletter

    \def\CT@@do@color{
      \global\let\CT@do@color\relax
            \@tempdima\wd\z@
            \advance\@tempdima\@tempdimb
            \advance\@tempdima\@tempdimc
    \advance\@tempdimb\tabcolsep
    \advance\@tempdimc\tabcolsep
    \advance\@tempdima2\tabcolsep
            \kern-\@tempdimb
            \leaders\vrule

                    \hskip\@tempdima\@plus  1fill
            \kern-\@tempdimc
            \hskip-\wd\z@ \@plus -1fill }
\makeatother

\renewcommand{\arraystretch}{1.2}

\date{\today}

\usepackage{xcolor}

\definecolor{Rayne}{HTML}{0090d9}

\newcommand{\ax}{{\rm a}}
\newcommand{\de}{{\rm de}}

\newcommand{\fa}{f_{\de}}
\newcommand{\rhoeff}{\rho_{\rm eff}}
\newcommand{\weff}{w_{\rm eff}}

\newcommand{\wde}{w_{\de}}

\newcommand{\rd}{r_{\rm d}}
\newcommand{\lgm}{{\rm lg}(m_\ax)}
\newcommand{\DMeff}{D_{M}({0.8})}
\newcommand{\Dstar}{D_{M}(z_*)}
\newcommand{\clpp}{C_\ell^{\phi\phi}}
\newcommand{\nolowlEE}{{\rm nolowE}}
\begin{document}
\title{Phantom Mirage from Axion Dark Energy}

\author{Rayne Liu}
\affiliation{Kavli Institute for Cosmological Physics, University of Chicago, Chicago, IL 60637, USA}

\author{Yijie Zhu}
\affiliation{Department of Physics and Astronomy, Stony Brook University, Stony Brook, NY 11794, USA}

\author{Wayne Hu}
\affiliation{Kavli Institute for Cosmological Physics, University of Chicago, Chicago, IL 60637, USA}

\author{Vivian Miranda}
\affiliation{C. N. Yang Institute for Theoretical Physics, Stony Brook University, Stony Brook, NY 11794, USA}

\begin{abstract}
Supernova (SN) and baryon acoustic oscillation (BAO) distance measures have recently provided hints that the dark energy is not only dynamical but apparently evolves from normal to phantom dark energy between redshifts $0<z<1$.  A normal axion dark energy component in the mass range just below the Hubble scale can mimic a phantom component by appearing as dark energy at $z=1$ and dark matter at $z=0$, raising the possibility of a phantom mirage.  We show that there is a wide range of axion dark energy contributions that can resolve the SN-BAO tension as well as thawing quintessence does, leaving BAO tension with the cosmic microwave background (CMB) for the distance measures from $z\sim 1$ to recombination to be resolved at high redshifts. With axions, raising the optical depth to reionization to $\tau \approx 0.1$ works essentially as well as $w_0-w_a$ phantom dark energy for all but the lowE CMB data, with a remaining $\Delta\chi^2\sim -16$
compared with $\Lambda$CDM, whereas a small spatial curvature of $\Omega_K \sim 0.003$ 
can  largely relax the full SN-BAO-CMB tension with a total $\Delta\chi^2 \sim -12$. 
\end{abstract}
\maketitle

\section{Introduction}

Recent distance-redshift measurements of Supernova (SN) Type Ia \cite{DES:2024jxu} and baryon acoustic oscillations (BAO) \cite{DESI:2025zgx,DESI:2025ejh} in conjunction with the cosmic microwave background (CMB) \cite{Planck:2018vyg} data have raised the possibility that the dark energy, which accelerates the expansion, is not only dynamical, but may also be best fit as a component that evolves from normal matter to matter that at least apparently if considered on its own violates the null energy condition, dubbed phantom dark energy \cite{DESI:2025fii,Ye:2024ywg,Jia:2025poj,Goldstein:2025epp,
Chaudhary:2025bfs,RoyChoudhury:2025iis,Kou:2025yfr,Goh:2025upc,Gomez-Valent:2025mfl,Brax:2025ahm,Braglia:2025gdo,Silva:2025twg,Wang:2025znm, Mishra:2025goj,Lee:2025rmg,Gialamas:2025pwv,Ozulker:2025ehg,Keeley:2025rlg,Cline:2025sbt,deSouza:2025rhv,Dinda:2025iaq,DESI:2025wyn,Berghaus:2024kra,Reboucas:2024smm,Loverde:2024nfi}.

Such an exotic component suggests a dark energy equation of state parameter $w_\de$ that evolves from $w_\de<-1$ to $w_\de>-1$ and places strong conditions on physical candidates in order not to exhibit ghost and gradient instability, which typically requires fundamental modifications to gravitational or dark sector forces
\cite{Vikman:2004dc,Hu:2004kh,Creminelli:2008wc,Fang:2008sn,Gubitosi:2012hu,Bloomfield:2012ff}, see also 
\cite{Oliveira:2025uye,Pullisseri:2025ran,Wolf:2024stt,Wolf:2025jed,Wolf:2025acj,Pourtsidou:2025sdd,Tsujikawa:2025wca,Yao:2025wlx,Guedezounme:2025wav,Hogas:2025ahb,Paliathanasis:2025hjw,Andriot:2025los,You:2025uon,Pan:2025qwy,Silva:2025hxw,SanchezLopez:2025uzw,Baryakhtar:2024rky,Bottaro:2024pcb,Costa:2025kwt} for recent assessments).
On the other hand, the well-established agreement with high redshift predictions in $\Lambda$CDM from the CMB indicates that these drastic modifications have the curious property that the strong deviations in the normal and phantom phases must nearly cancel each other so as to have similar endpoints to the expansion history as $\Lambda$CDM  \cite{Hu:2004kh,Linder:2007ka}, implying that one or the other may be a mirage.  

It is therefore important to examine other ways in which this implied expansion history can occur without ever requiring a phantom dark energy component: that phantom dark energy is a mirage. One well-studied possibility
is a coupling between the dark matter and dark energy, or the so-called coupled quintessence, where the dark energy transfers some of its energy density to the dark matter so that if we analyze its expansion history assuming decoupled cold dark matter, we would infer a phantom equation of state for the effective dark energy density \cite{Amendola:1999er,Das:2005yj,Copeland:2006wr,Khoury:2025txd,Bedroya:2025fwh}.  Similar to the phantom models, these models still typically introduce new forces between dark matter and dark energy in order to mediate the coupling.  

In this work, we study the ability of axions to provide a similar phantom mirage without introducing any additional couplings in the dark sector, requiring only an additional cosmological constant or normal quintessence.   It is well known that a light scalar field acts as a contribution to the cosmological constant  when it is frozen on its potential by Hubble friction, and once released acts instead as dark matter. The QCD axion, for example, is a leading candidate to compose of all of the dark matter \cite{Preskill:1982cy,Abbott:1982af,Dine:1982ah}.  Moreover, axions as all of the dark energy provide a technically natural solution through their shift symmetry as to why their mass can stay of order the very small $m_\ax \sim H_0 \sim 10^{-33}$\,eV required \cite{Carroll:1998zi,Shajib:2025tpd}.  If instead the mass is only somewhat larger than this Hubble scale value, then the transition between dark energy and dark matter behavior can occur between redshifts $0 < z< 1$ and provide a phantom mirage for SN and BAO distance measures, and allow non-dark-energy explanations for the BAO-CMB tension in $\Lambda$CDM to accommodate all three types of data.

In Sec.\ \ref{sec:methods}, we give the SN, BAO and CMB datasets, axion dark energy parameters and statistical methods employed.  
In Sec.\ \ref{sec:phantom_mirage}, we discuss why axion dark energy can produce the mirage of phantom dark energy and even phantom crossing.  
We present the comparison with the data in Sec.\ \ref{sec:results} and discuss these results in \ref{sec:discussion}.   In Appendix \ref{sec:emulator_training}, we provide details on our fast and efficient emulation approach for predicting axion model observables.

Throughout we employ units where $\hbar=c=1$ and define $\lgm= \log_{10}(m_\ax/1{\rm eV})$ but explicitly write $\log_{10}$ elsewhere.

\section{Datasets and Methods}
\label{sec:methods}

We study the ability of axions to relax the tension between the following baseline data sets (see Appendix \ref{sec:emulator_training} for calculational details and definitions): 
\begin{itemize}

    \item CMB: Planck 2018 \cite{Planck:2018vyg} PR3 lowT, lowE, and plik TTTEEE lite likelihoods, Planck PR4 lensing \cite{Carron:2022eyg} and ACT DR6 lensing baseline data \cite{ACT:2023dou}.  When comparing datasets to models we take the  $\Lambda$CDM Planck 2018 TTTEEE+lowT+lowE+lensing mean parameters \cite{Planck:2018vyg} (their Tab.~2) as the fiducial model, denoted ``fid" for shorthand.
    
    \item BAO: DESI DR2 \cite{DESI:2025zgx} distance measures, using their Tab.~4 for $D_V/\rd$ for the lowest redshift bin and $D_M/\rd$, $D_H/\rd$, and their cross-correlation $r_{M,H}$ for all other bins.  When plotting any observable in units of the predictions in the fiducial model we denote, e.g.\  
    $D_M(z) \, [{\rm fid}]= D_M(z)/D_{M,{\rm fid}}(z)$.
    
    \item SN: DESY5 supernovae distance modulus $\mu$ \cite{DES:2024jxu}.  The likelihood analysis uses the unbinned data but for plotting purposes we bin in redshift using inverse covariance weights and compare to the fiducial model after optimizing the unkwown absolute magnitude. 
    
\end{itemize}
We choose the most recent BAO and SN data sets as they exhibit the strongest evidence for phantom dark energy and hence the largest challenge to resolve with axions.

We call the best fit $\Lambda$CDM model to these datasets the ``baseline $\Lambda$CDM" model, which is distinct from the Planck 2018 fiducial model due to the compromises in parameter values to fit these datasets.

In addition to the usual 6 $\Lambda$CDM parameters (initial curvature spectrum amplitude $A_s$, spectral tilt $n_s$, angular size of the sound horizon $\theta_s$, baryon density $\Omega_b h^2$, cold dark matter density $\Omega_c h^2$ and optical depth to reionization $\tau$)
with their usual  uninformative priors \cite{Planck:2018vyg}, we add the axion mass $m_\ax$ with a flat, range bound prior of $-35 < \lgm < -31.5$ and their energy density $\Omega_\ax h^2$ with flat uninformative priors.  In an extension  to the baseline analysis to help resolve CMB+BAO tension, we also consider the spatial curvature $\Omega_K$ with flat uninformative priors.

\begin{figure}
        \centering
              \includegraphics[width=1\linewidth]{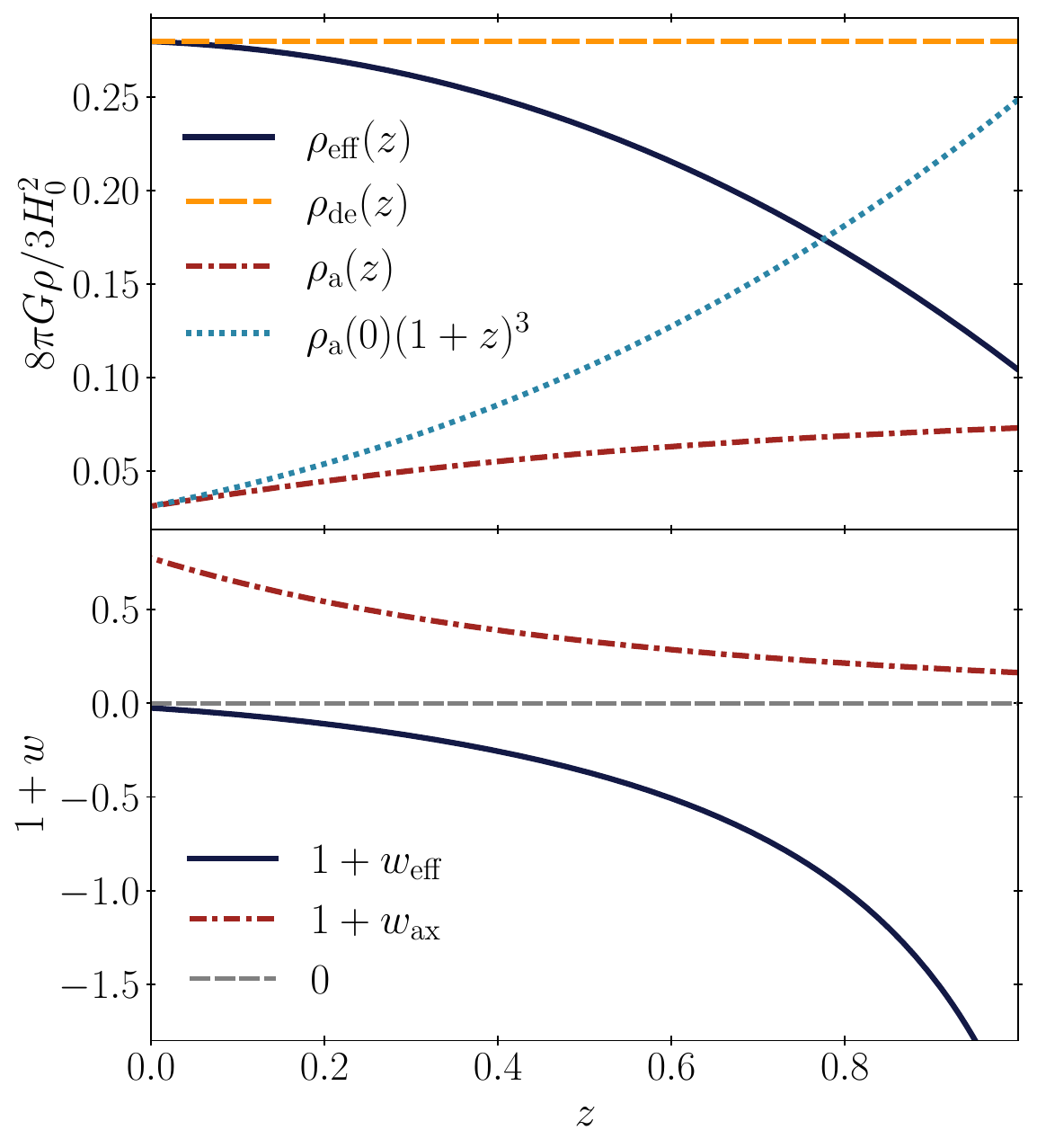}
        \caption{Phantom mirage mechanism.  Top panel: true energy densities of the axion $\rho_\ax$ and dark energy $\rho_\de$ compared with the effective dark energy $\rho_{\rm eff}$ that would be assumed by analyzing this case as CDM and dark energy. 
        Bottom panel: the inferred effective equation of state $w_{\rm eff}<-1$ at high redshift despite $w_\ax>-1$ approaching a CDM-like $w_\ax \sim 0$ at $z=0$. Here $\lgm =-32.5, \fa = 0.1.$
       }
\label{fig:pedagogical}
    \end{figure}

For the parameter estimation, we employ a fast and efficient  emulator of the  axion Einstein-Boltzmann code \textsc{AxiECAMB}\footnote{\url{https://github.com/Ra-yne/AxiECAMB}} \cite{Liu:2024yne} as detailed in  Appendix \ref{sec:emulator_training}.

In addition to Markov Chain Monte Carlo parameter posteriors using \textsc{Cobaya}\footnote{\url{https://github.com/CobayaSampler/cobaya}}, we show the profile likelihood as $\Delta\chi^2$ at maximum posterior using the
\textsc{Cocoa}\footnote{\url{https://github.com/CosmoLike/cocoa}} implementation of the  \textsc{Procoli} algorithm \cite{Karwal:2024qpt}.   Specifically  for each $\lgm$, we maximize the posterior $\ln {\cal P}$ over the other parameters, remove the prior $\ln {\cal L} = \ln {\cal P}-\ln \Pi$ and display the change in $-2\ln {\cal L}$ over that of the  baseline $\Lambda$CDM model. To penalize models with extra complexity for the same goodness of fit, we employ the 
Akaike Information Criterion 
$\Delta {\rm AIC} = 2 \Delta p - 2\Delta\ln{\cal L}$ where $\Delta p$ is the additional number of fitted parameters. 

\section{Phantom Mirage}
\label{sec:phantom_mirage}

Distance measures from SN, BAO and the CMB depend only on the expansion history and not the theoretical division of the dark sector into dark matter and dark energy components.  
An incorrect assumption about the dark matter can lead to an incorrect inference of phantom dark energy, which we call a phantom mirage.

A well known example of the phantom mirage is provided by coupled quintessence \cite{Amendola:1999er,Das:2005yj,Copeland:2006wr}.
In this case, coupling between the dark energy and dark matter transfers energy from the former to the latter.   When analyzed under the assumption of non-interacting cold dark matter, 
the remaining effective dark energy gives the mirage of a phantom component with $w<-1$.

Axions with $m_\ax \gtrsim H_0$ provide another example, but one that does not require  an additional coupling.  At early times when $H\gg H_0$, Hubble friction prevents the axion field from rolling in its potential causing its energy density to behave as an addition to the cosmological constant whereas at late times the field oscillates around the quadratic minimum and behaves as  dark matter.   A key difference with coupled quintessence is that for the axion phantom mirage, the transition from dark energy to dark matter does not involve coupling between the two but rather a separate component that mimics their respective behaviors at different redshifts.

To see that this produces the same expansion history as phantom dark energy and cold dark matter, define the effective dark energy density as that which would be inferred by considering axions as a contribution to CDM at the present.   Given a true dark energy density $\rho_{\de}$, we would instead infer an effective dark energy
of
\begin{equation}\label{eqn:rhoeff}
\rhoeff(z) \equiv \rho_{{\de}}(z) + \rho_\ax(z) - \rho_\ax(0) (1+z)^{3}
\end{equation}
so that the expansion history remains the same.  The equation of state of the effective dark energy is defined by 
\begin{align}\label{eqn:weff}
1+\weff &= \frac{1}{3} \frac{d\ln \rhoeff}{d\ln (1+z)} \\
& =
\frac{ (1+w_{\de}) \rho_{\de} + (1+w_\ax)\rho_\ax -  \rho_\ax(0) (1+z)^{3} }{\rhoeff} , \nonumber
\end{align}
where $w_{\de}$ and $w_\ax$ are the true dark energy and axion equations of state respectively.
At $z=0$, $\weff \approx \wde$, but for higher $z$ when $w_\ax < 0$ and approaches $-1$,  while  $\rhoeff>0$, the negative contribution from the incorrect matterlike assumption for axions drives
$\weff<-1$ which implies the mirage of a phantom.  We can interpret the negative contribution to $\rhoeff$ as reducing the high $z$ dark energy, which makes the effective dark energy grow as the universe expands.   If $w_\de>-1$, then this also implies a phantom crossing where $\rhoeff$ switches from growing to decaying with the expansion.  However, we  start with the simplest assumption that the true dark energy is a cosmological constant where $w_\de=-1$, and hence the case where the effective dark energy is always a phantom, at least when time averaged over axion oscillations.

\begin{figure}
    \centering   \includegraphics[width=1\linewidth]{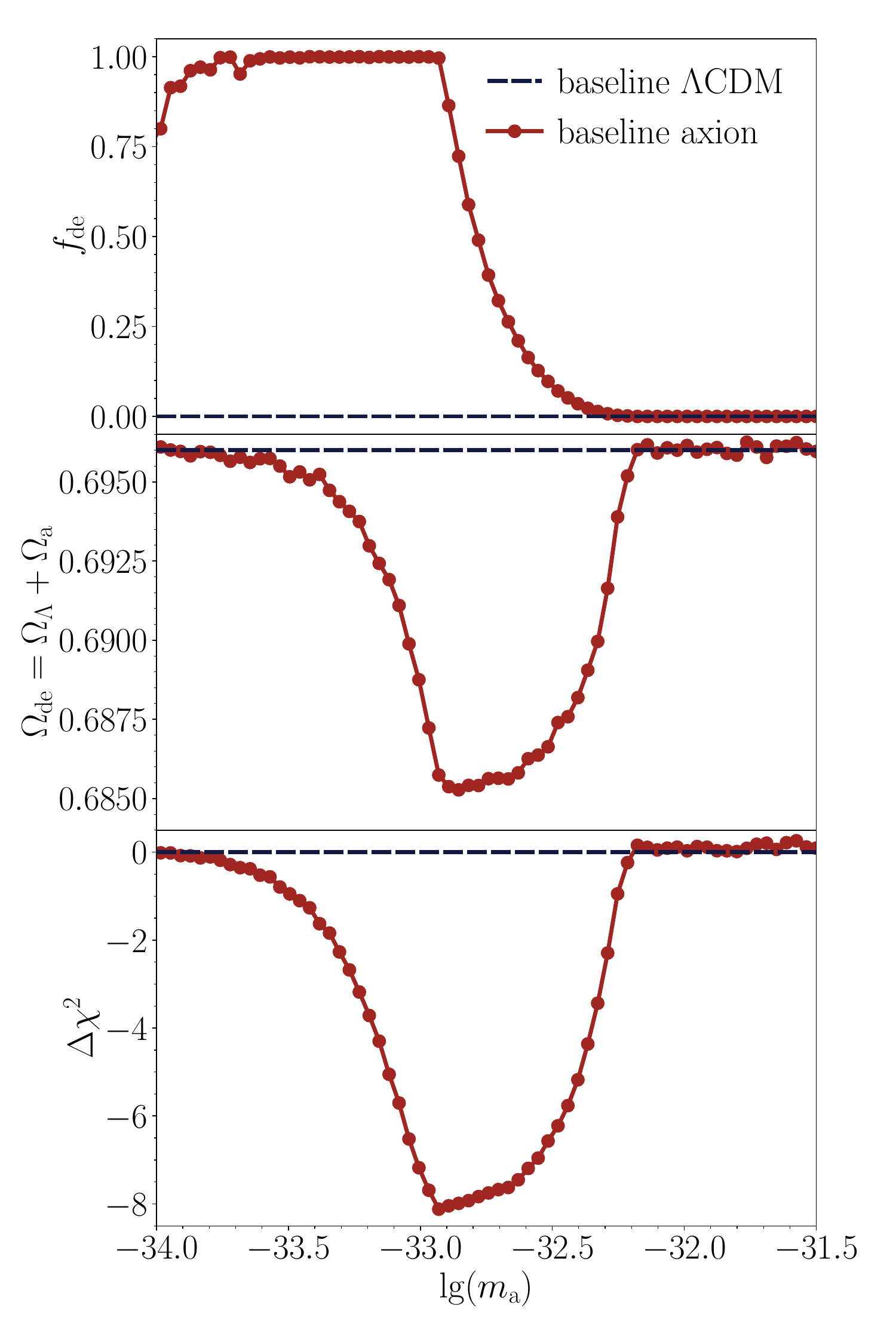}    \caption{Likelihood profile as a function of axion mass or $\Delta\chi^2$ relative to the baseline $\Lambda$CDM model for the baseline SN+BAO+CMB analysis (lower panel). Similar improvements over $\Lambda$CDM occur in the axion phantom mirage range $-33.5 \lesssim \lgm \lesssim -32.3$ but with very different axion contributions to the dark energy $f_\de = \Omega_\ax/\Omega_\de$ (upper panel).  For much smaller masses all $f_\de$ cases behave as $\Lambda$CDM with the same total $\Omega_\de = \Omega_\Lambda+\Omega_\ax$ (middle panel) and for much higher masses the profile value for $f_\de$ is small and also predicts $\Lambda$CDM observables up to the accuracy of  the emulator discussed in Appendix \ref{sec:emulator_training}.}\label{fig:profile_fixedmass}
\end{figure}

\begin{table*}[t]
\centering
\renewcommand{\arraystretch}{1.2} 
\begin{tabular}{c| cc|c|cccc}
name  & $\lgm$ & $f_\de$ & extension & $\Delta \chi^2$ &  $\Delta \chi^2_{\rm no lowE}$ & & \\
\hline
baseline $\Lambda$CDM  & -- & 
$0$
& 

--
& 
$0$ 
&
$0$ 
&
&
\\
$w_0-w_a$ & -- & 
$0$
& 
$w_0=-0.76,w_a=-0.815$
& 
$-19.4$
&
$-18.3$
&
&
\\
\hline
baseline & $-32.9$ & 
$0.986$
& 

--
& 
$-8.1$
&
$-9.2$
&
&
\\
& $-32.5$ & 
$0.086$
& 
--
& 
$-6.5$
&
$-7.5$
&
&
\\
\hline
no lowE & $-32.9$ & 
$0.999$
& 

$\tau=0.1$
& 
--
& 
$-15.8$
&
&
\\
 & $-32.5$ & 
$0.116$
& 
$\tau=0.102$
& 
--
& 
$-14.5$
&
&
\\
\hline
curvature & $-32.9$ & 
$0.849$
& 

$\Omega_K=0.0027$
& 
$-12.5$
&
$-12.4$
&
&
\\
 & $-32.5$ & 
$0.09$
& 
$\Omega_K=0.0029$
& 
$-11.3$
&
$-11.1$
&
&
\\

\end{tabular}
\caption{Model parameters in the baseline and extended analyses with high optical depth $\tau$ or spatial curvature $\Omega_K$.   All $\Delta\chi^2$ values are relative to the baseline $\Lambda$CDM  model.  For axion cases the best fit $\lgm=-32.9$ and the high mass end of the axion phantom mirage $\lgm=-32.5$ are shown and generally perform comparably despite the very different dark energy fraction $f_\de$.  High $\tau$ models drop the lowE Planck 2018 data and fit the rest of the SN+BAO+CMB data almost as well as $w_0-w_a$. 
}
\label{tab:dchi2_table}
\end{table*}

We illustrate this behavior in Fig. \ref{fig:pedagogical} (top panel) with an example where $\lgm =-32.5 $, close to the mass where the  equation of state at the present first reaches $w_\ax(0)=0$.
More precisely, with the specific choice of $h=0.674$, $\Omega_\de=0.311$ 
and $f_\ax=0.1$ employed here,  $w_{\ax}(0)=0$ at
$\lgm = -32.45$.
Notice that by $z=1$, the axions have already reached the frozen regime and an extrapolation as matter would overestimate the matter density.  The effective energy density then decreases at high redshift implying a phantom equation of state $\weff<-1$ even though $w_\ax >-1$ and $w_\de=-1$, see Fig.~\ref{fig:pedagogical} (bottom panel).
Phrased alternatively, the expansion history mimics a low $\Omega_m$ model at the high redshifts $z\gtrsim 0.5$ relevant for BAO and a high $\Omega_m$ for the low redshifts $z\lesssim 0.5$ relevant for SN.   The axion phantom mirage can therefore reduce the tension between BAO and SN without having any phantom component.  Furthermore, by having only a small fraction $f_\de$ of dark energy in axions but with a larger change in their equation of state, the axion phantom mirage can produce a more rapid change in $\weff$  than a model with just a single component of thawing quintessence.

\section{Axion landscape}
\label{sec:results}

\begin{figure}
    \centering
    \includegraphics[width=1\linewidth]{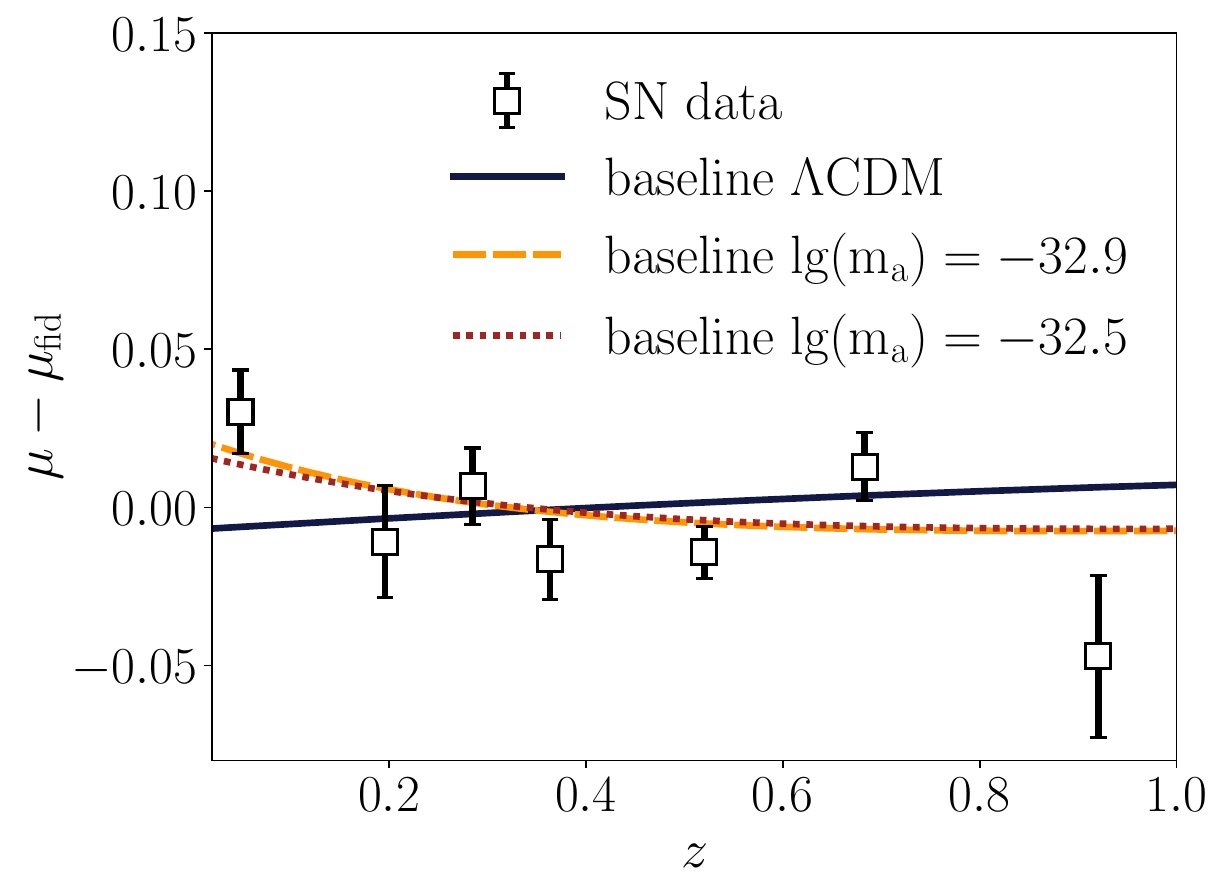}
    \caption{SN data compared with models in the baseline analysis.  The baseline $\Lambda$CDM  model fails to fit the relative magnitudes of the lowest redshift SN vs higher redshifts whereas both axion cases 
    $\lgm=-32.9$ ($f_\de\approx 1$) and $\lgm=32.5$ ($f_\de \approx 0.09$) capture the upturn.                              }
\label{fig:SN_examplemasses}
\end{figure}

\begin{figure*}
    \centering
    \includegraphics[width=1\linewidth]{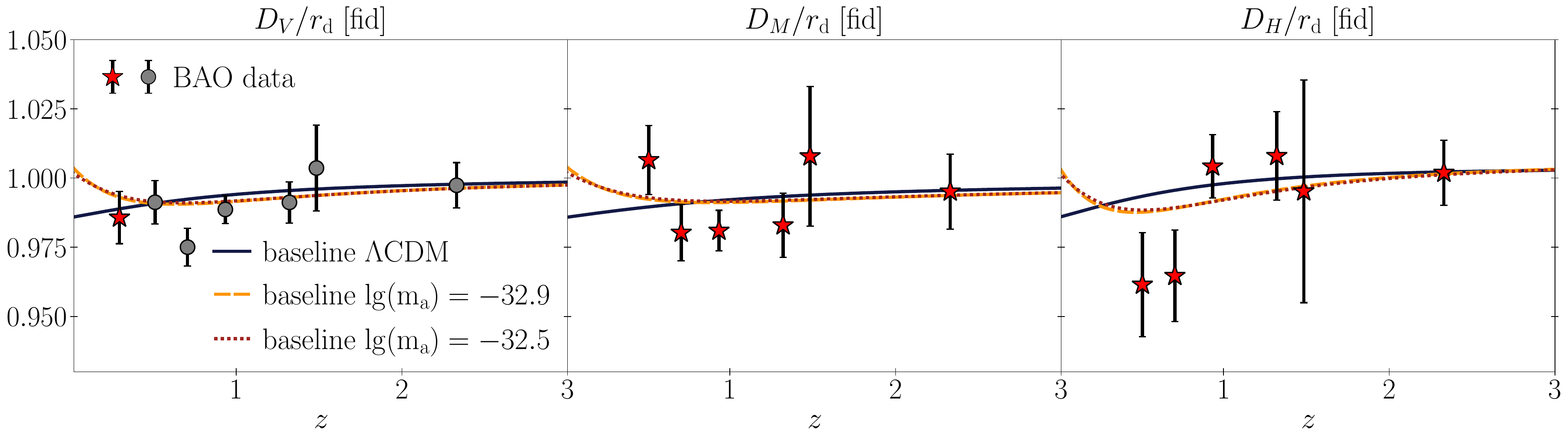}
    \caption{BAO data compared with models in the baseline analysis.
    While the baseline $\Lambda$CDM  lacks the low $z$ upturn of the axion cases, all three fit the high-$z$ BAO data comparably well. Here and in other figures ``[fid]" stands for in units of the same quantity in the fiducial model.   Starred datapoints represent those which are used in the likelihood whereas $D_V$ for all but the lowest redshift bin  is redundant with $D_M$ and $D_H$. }
    \label{fig:BAO_examplemasses}
\end{figure*}

\subsection{Baseline analysis}

We begin our analysis with the simplest scenario where axions are the only addition to the standard $\Lambda$CDM paradigm and all data sets are taken at face value, which we call the baseline analysis.

In Fig.~\ref{fig:profile_fixedmass} (bottom panel), we show the $\Delta\chi^2$ profile in the axion mass $m_\ax$ relative to the baseline $\Lambda$CDM model, as well as the values of the axion fraction of the dark energy $f_\de$ (top panel) and total dark energy $\Omega_\de=\Omega_\Lambda+\Omega_\ax$  (middle panel) that correspond to these profile models.  
Note that for $\lgm \lesssim -33.5$,
despite the apparent changes in the value of $f_\de$ for different $m_\ax$,  all $f_\de$ for the given $\Omega_\de$ fit essentially equally well, with the changes in $\Delta\chi^2$ being much less than unity as the profile shows.  
This is because in this  $m_\ax \ll H_0$ limit, the axion field is still frozen on its potential at the present.

In the range $-33.5 \lesssim \lgm  \lesssim -32.3$,
cases with finite $1>f_\de>0$ fit the data better than $\Lambda$CDM.
The low mass end of this regime represents the standard thawing quintessence case $f_\de \sim 1$ where all of the dark energy is in a single scalar field component.    Interestingly, the higher masses in this range provide nearly as good a fit but with a steeply decreasing $f_\de$.
 This extension to higher masses and lower fractions displays the axion phantom mirage discussed in Sec.~\ref{sec:phantom_mirage} where the axion behaves increasingly like dark matter at $z=0$.

Given the nearly equally good fits across this axion phantom mass range, we select two representative cases with very different $f_\de$ and examine the origin of the preferences in the data.  The relevant parameters of these models are given in Tab.~\ref{tab:dchi2_table}.  The first is the global minimum at $\lgm =-32.9$ and $f_\de \approx 1$, with
$\Delta\chi^2= -8.1$ compared to $\Lambda$CDM and the second is
$\lgm =-32.5$ and $f_\de \approx 0.09$, with
$\Delta\chi^2=-6.5$.  Both are comparable to the improvements in the calibrated thawing quintessence models where $w_\de(z) =w_0 - 1.58(1+w_0)z/(1+z)$
studied in Ref.~\cite{DESI:2025fii}. 
Note that for each mass in our phantom mass range considered separately as in the profile, there is one additional parameter, the same as in thawing quintessence.  Thus for example compared to $\Lambda$CDM, the $\lg(m_\ax)=-32.5$ case has $\Delta {\rm AIC} = -4.5$.

In particular both models provide comparable fits to the SN data,  shown in Fig.~\ref{fig:SN_examplemasses}, especially the upturn at the lowest redshift bin, despite their different $f_\de$.    This is a consequence of the larger mass producing a much sharper change in distances for a given $f_\de$ due to its equation of state quickly approaching a matter like value of $w_\ax\sim 0$. 
Baseline $\Lambda$CDM has the opposite trends with redshift given that it is optimized to SN+BAO+CMB and does not fit the SN data alone well.

In Fig \ref{fig:BAO_examplemasses}, we show the comparison to BAO distance measures.  Again the two cases fit BAO comparably well but notice that the fits are also comparable to baseline $\Lambda$CDM, especially for $z>0.5$ where the data are the most constraining.    In particular around $z\sim 0.8$ all three models overpredict $\DMeff/\rd$ by around a percent.  
This is a consequence of the high redshift tension between BAO and CMB given the calibration of $\rd$ provided by the latter and its inference of $\Dstar$, the distance to recombination at $z_*$, through the precise measurement of $\theta_*$, the angular size of the sound horizon.  Axions and more generally thawing dark energy models cannot efficiently change $\Dstar-\DMeff=\int_{0.8}^{z_*} dz/H(z)$ in a flat universe.

\begin{figure}
    \centering
    \includegraphics[width=1\linewidth]{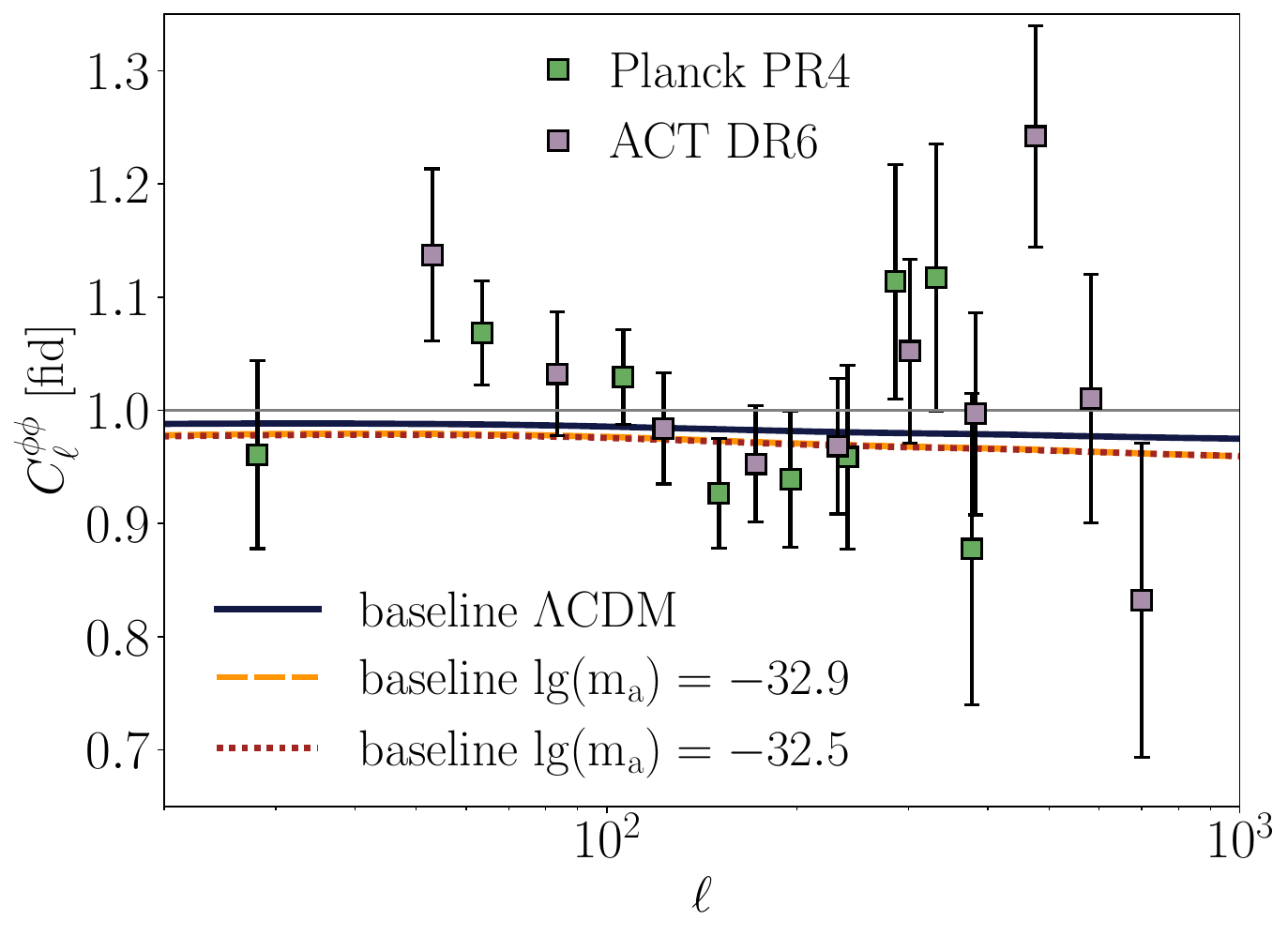}
    \caption{CMB lensing data compared with models in the baseline analysis. The baseline $\Lambda$CDM  and axion cases all reduce the lensing amplitude so as to fit the BAO data reflecting tension in the BAO+CMB data.   }
    \label{fig:Clphiphi_examplemasses}
\end{figure}

The CMB calibration of $\rd$ is increasingly impacted by CMB lensing measurements as ground based data improve.  In Fig.~\ref{fig:Clphiphi_examplemasses}, we show the comparison of the models to the data for the CMB lensing power spectrum $\clpp$ relative to the fiducial model.  
Notice that with SN+BAO+CMB, all three best fit curves further underpredict lensing relative to the fiducial model and are a slightly worse fit to the data.  
This is again the consequence of the compromise between the BAO and CMB data. 
In fact, in the fiducial model the lensing amplitude is itself slightly low compared with additional SPT lensing data that is not included here which would only strengthen this conclusion (see e.g.~\cite{SPT-3G:2024atg} for a recent $A_L$  rescaling type assessment).

\begin{figure}
    \centering
    \includegraphics[width=1\linewidth]{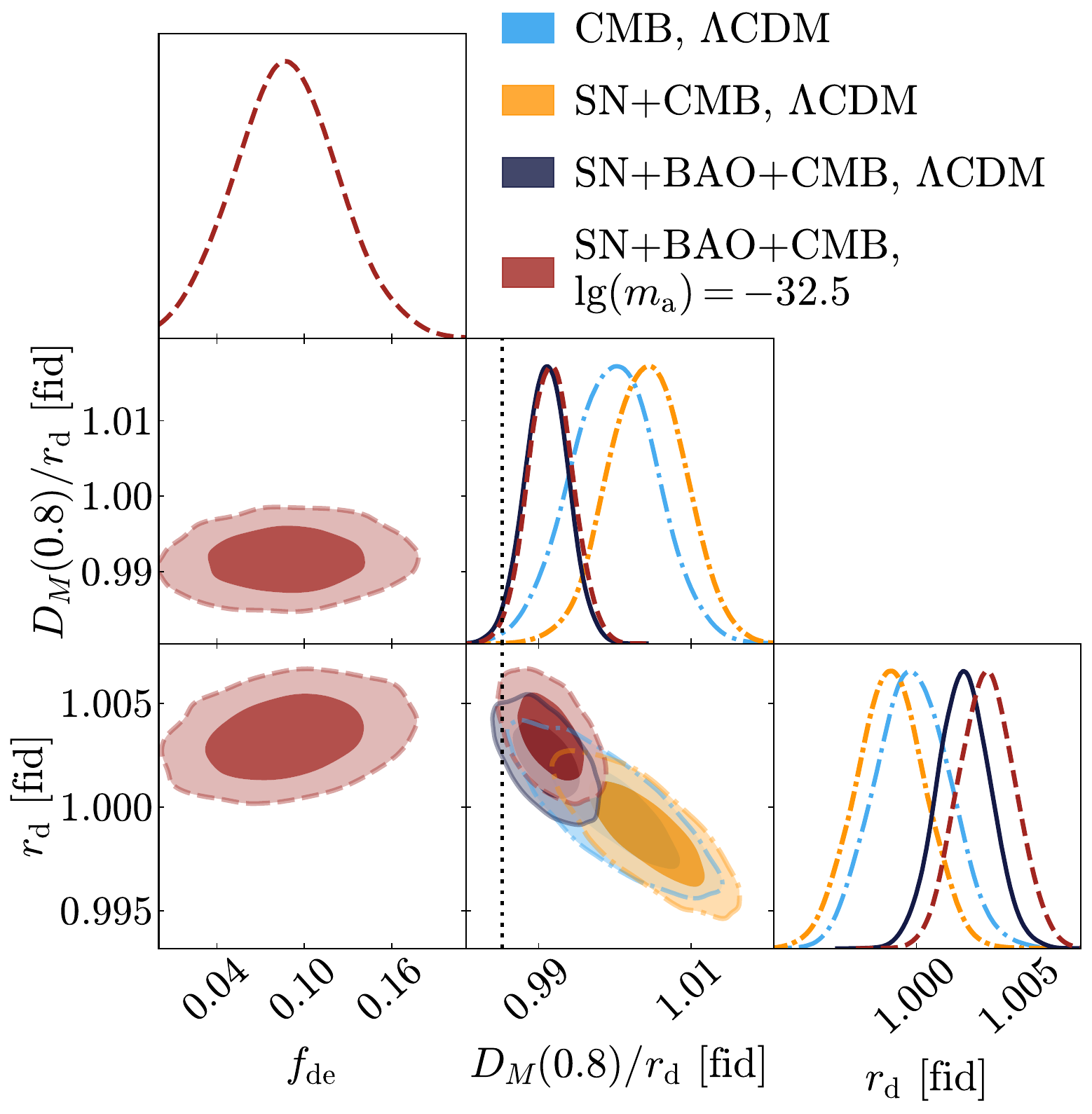}
    \caption{BAO+CMB tension parameters vs.~axion dark energy fraction in the baseline analysis (68\%, 95\% CL contours). BAO data only (vertical dotted line taken from Ref.~\cite{DESI:2025zgx} in their Fig. 6) prefer a lower value for $\DMeff/\rd$ than in the fiducial CMB only analysis and this tension is exacerbated by the inclusion of SN so that the baseline $\Lambda$CDM analysis of SN+BAO+CMB pushes constraints into the tail of the SN+CMB posterior.  The axion case with $\lgm =-32.5$ shows a 95\% exlcusion of $f_\de=0$ due to the better fit to the lowest $z$ SN but does not appreciably change the BAO+CMB tension between $z=0.8$ and recombination. }
\label{fig:posterior_miragemass_tension}
\end{figure}

On the CMB side, to fit the acoustic peaks and the amplitude of the lensing power spectrum, the cold dark matter $\Omega_c h^2$  must remain relatively high and $\rd$ relatively low. This parameter also controls the BAO vs.\ CMB distances $\Dstar-\DMeff$, with BAO measurements favoring the lower $\Omega_c h^2$.  The lack of freedom to adjust these relationships causes the tension between the BAO and CMB data.

We can see this directly in Fig.~\ref{fig:posterior_miragemass_tension}.   Here we show the posterior constraints for $f_\de, \DMeff/\rd,\rd$ for $\Lambda$CDM vs.~axions with $\lgm = -32.5$.   In $\Lambda$CDM, the tension between the CMB and BAO distance measures appears as a preference for high $\DMeff/\rd$ and low $\rd$ in the former vs.\ low $\DMeff/\rd$ in the latter.   Adding in SN to the CMB, while still consistent in $\Lambda$CDM, makes the tension worse.   This is the well known SN+BAO discrepancy in $\Omega_m$ and BAO+CMB shift in $H_0 \rd$ but phrased more directly in terms of quantities relevant to BAO and hence in a more  model independent fashion that is useful for dynamical dark energy as well.

When BAO are added to SN+CMB, $\DMeff/\rd$ and $\rd$ are pushed to the extremes of what is allowed by the latter, reflecting the tension in the data sets.
On the other hand the improvement from better fitting the SN is enough to place $f_\de=0$ outside the $95\%$ CL regions.

Thus while the axion phantom mirage provides a good fit to SN for a wide range of fractional contributions
to the dark energy  from $5\%$ to $100\%$, it cannot resolve all of the tension between SN+BAO+CMB but instead performs about as well as the calibrated thawing quintessence class.

\begin{figure}
    \centering
    \includegraphics[width=\linewidth]{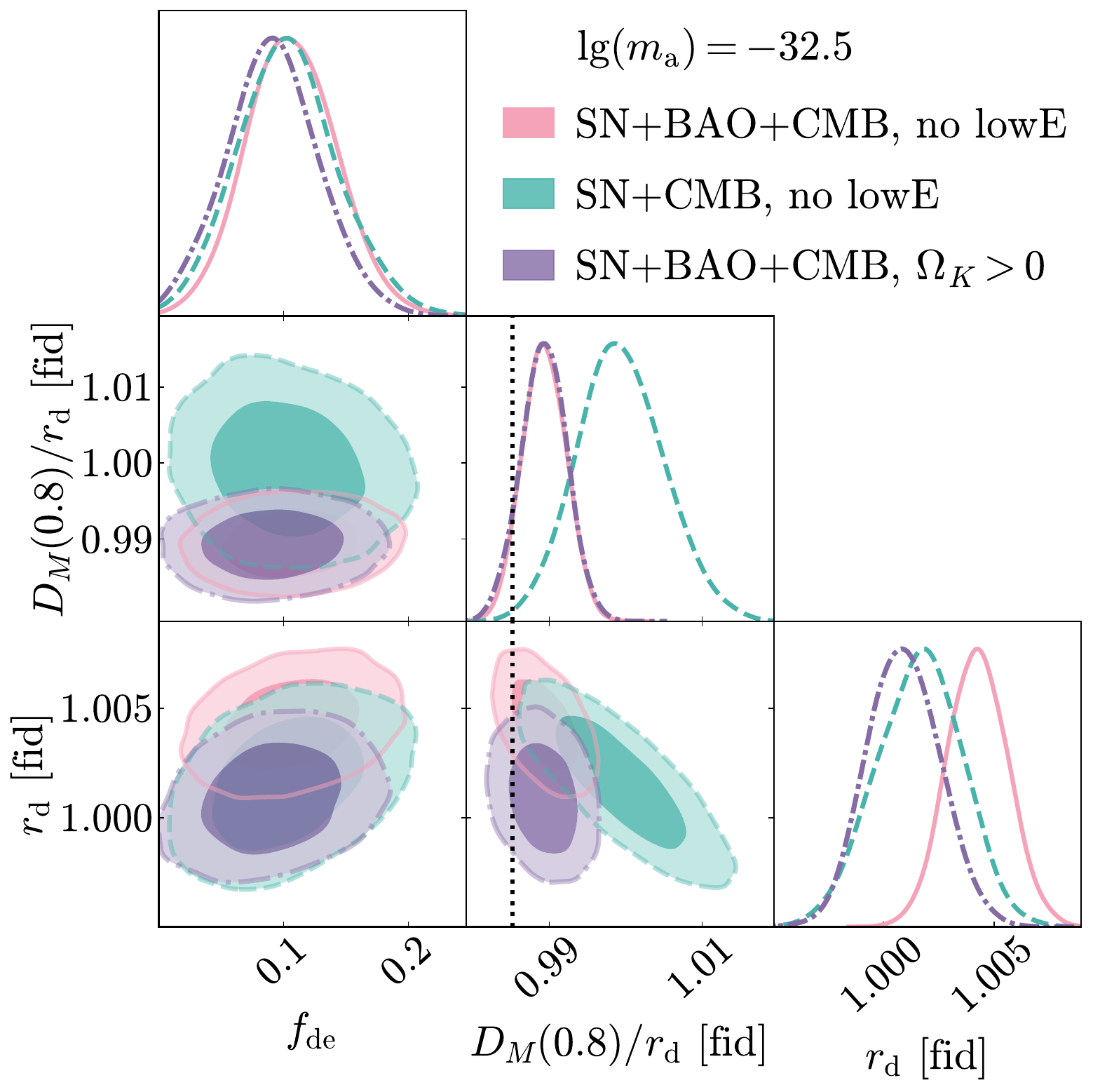}
     
    \caption{BAO+CMB tension parameters  vs.~axion dark energy fraction in the high-$z$ extension analyses.  Tension in the $\DMeff/\rd,\rd$ plane can be addressed by increasing $\tau$, here through dropping the lowE Planck 2018 data, or by adding spatial curvature.  
Vertical dotted line shows the  BAO-only preference for $D_M(0.8)/\rd$. 
    The former allows the CMB calibration of $\rd$ to shift whereas the latter changes the relation between $\DMeff$ and the distance to recombination.  Both provide better fits to SN+BAO+CMB than the baseline case and retain a preference for $f_\de>0$ from SN at this mass, $\lgm=-32.5$.
    }
    \label{fig:posterior_miragemass_resolution}
\end{figure}

\subsection{High-$z$ extensions}

We have seen that a wide range of axion contributions to the dark energy  fit the SN+BAO+CMB data as well as calibrated thawing quintessence but fail to address the BAO+CMB tension. 
This is because by construction such axions mainly affect distances to low redshift $z<0.5$, leaving the high redshift universe to resemble $\Lambda$CDM, especially in the distance between the best BAO constraints around $z \sim 0.8$ and recombination.

On the other hand, it is well known that the BAO+CMB tension alone can be resolved without dynamical dark energy by either changing the CMB calibration of $\rd$ or the  distance between $z\sim 0.8$ and recombination, but these solutions fail to fit SN data at much lower redshifts. When such solutions are combined with the axion phantom mirage, they jointly address the full SN+BAO+CMB tension.

As has been emphasized in Ref.~\cite{Jhaveri:2025neg} (see also \cite{Sailer:2025lxj,Giare:2023ejv,Loverde:2024nfi,Allali:2025wwi,Allali:2025yvp,Huang:2025xyf,Elbers:2025xvk}),
in $\Lambda$CDM the CMB+BAO tension rests on the inability to change the cold dark matter density $\Omega_c h^2$ given the measurements of the CMB lensing  once the optical depth to reionization $\tau$ is constrained by the Planck lowE polarization data.  In combination with Planck measurement  of the amplitude of the acoustic peaks, which is controlled by $A_s e^{-2\tau}$, the optical depth constraint prevents  the amplitude $A_s$ of the curvature power spectrum, from adjusting the lensing amplitude to fit the data and instead fixes $\Omega_c h^2$.   Even without CMB lensing reconstruction, lensing constraints from the smoothing of the acoustic peaks comparably limit this ability, which suggests that the tension is not due to unknown systematics in the lensing reconstruction. With $\Omega_c h^2$ fixed, $\Lambda$CDM no longer has the freedom to adjust $\Dstar-\DMeff$.

   These constraints on $\tau$ can be somewhat relaxed in extended reionization models \cite{Heinrich:2021ufa,Tan:2025obi}  or alternately by reducing the large scale primodial power spectrum from inflation \cite{Obied:2018qdr,Jhaveri:2025neg}.  There may also be unknown systematics in this challenging measurement \cite{Delouis:2019bub,CLASS:2025khf}.  Here we remain agnostic about the physical origin of this solution and simply test it by removing the lowE likelihood contribution in the data following 
\cite{Sailer:2025lxj,Jhaveri:2025neg}.
   Doing so, the remaining CMB data in $\Lambda$CDM give a lower $\Omega_c h^2=0.1187\pm 0.0014$  vs.~$0.12$ in the fiducial model and a better fit to TTTEEE in particular.

\begin{figure}
    \centering
    \includegraphics[width=1\linewidth]{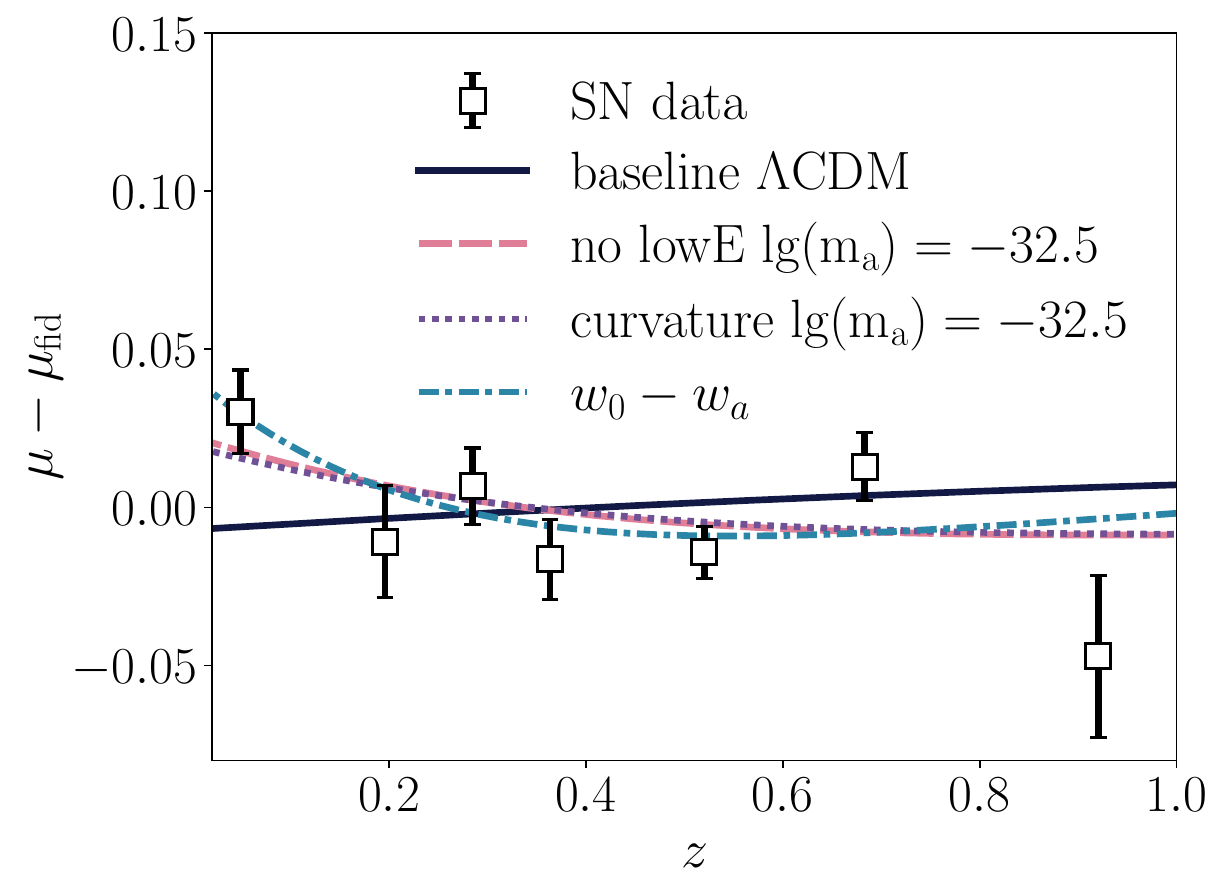}
    \caption{SN data compared with models in the extended analyses, no lowE and curvature at $\lgm=-32.5$, vs $w_0-w_a$. The baseline $\Lambda$CDM case is included for reference. All three models fit the SN data nearly equally as well. }
    \label{fig:SN_extensions}
\end{figure}

\begin{figure*}
    \centering    \includegraphics[width=1\linewidth]{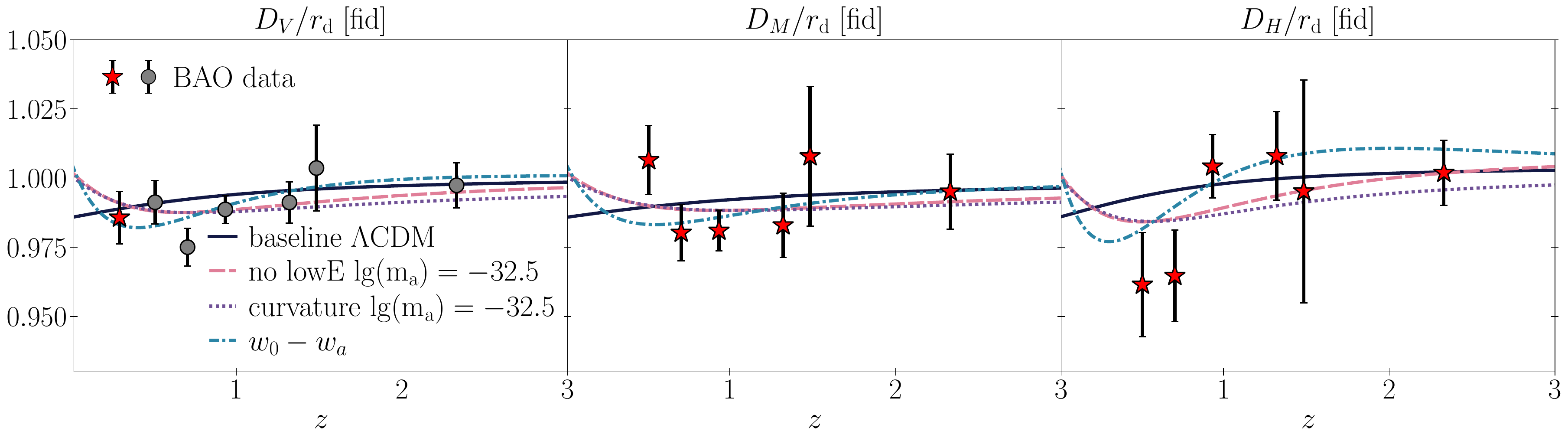}
    \caption{BAO data compared with models in the extended analyses, no lowE and curvature at $\lgm=-32.5$, vs $w_0-w_a$.  The no lowE case fits nearly as well as $w_0-w_a$ whereas the curvature case fits slightly worse at high-$z$ but still better than the baseline $\Lambda$CDM model. }
    \label{fig:BAO_extensions}
\end{figure*}

\begin{figure}
    \centering
    \includegraphics[width=1\linewidth]{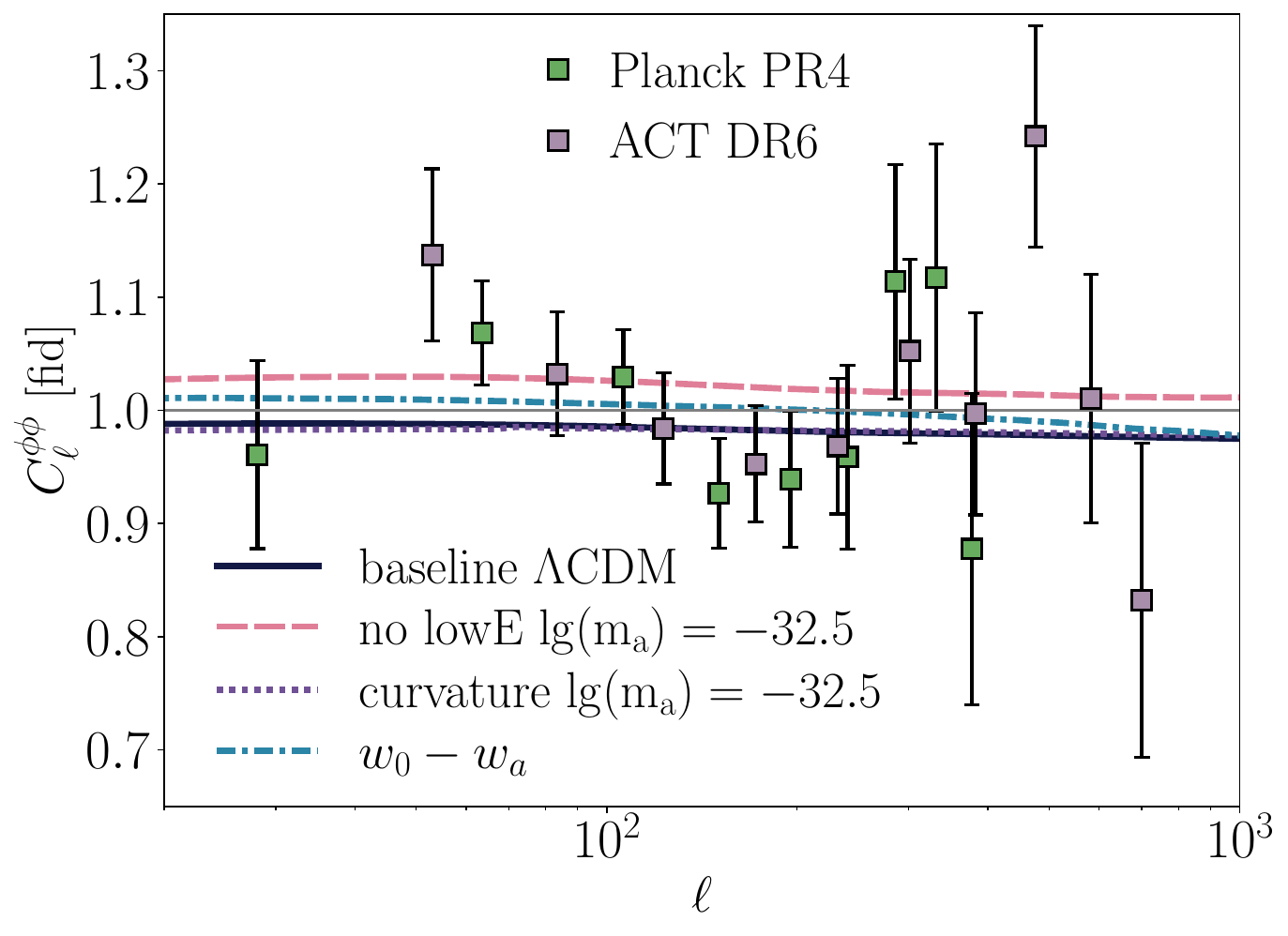}
\caption{CMB lensing data compared with models in the extended analyses, no lowE and curvature at $\lgm=-32.5$, vs $w_0-w_a$.
     The no lowE case fits as well as $w_0-w_a$ whereas the curvature case fits as well as the baseline $\Lambda$CDM model.
}
    \label{fig:Clphiphi_phantomextensions}
\end{figure}

\begin{figure}
    \centering
    \includegraphics[width=1\linewidth]{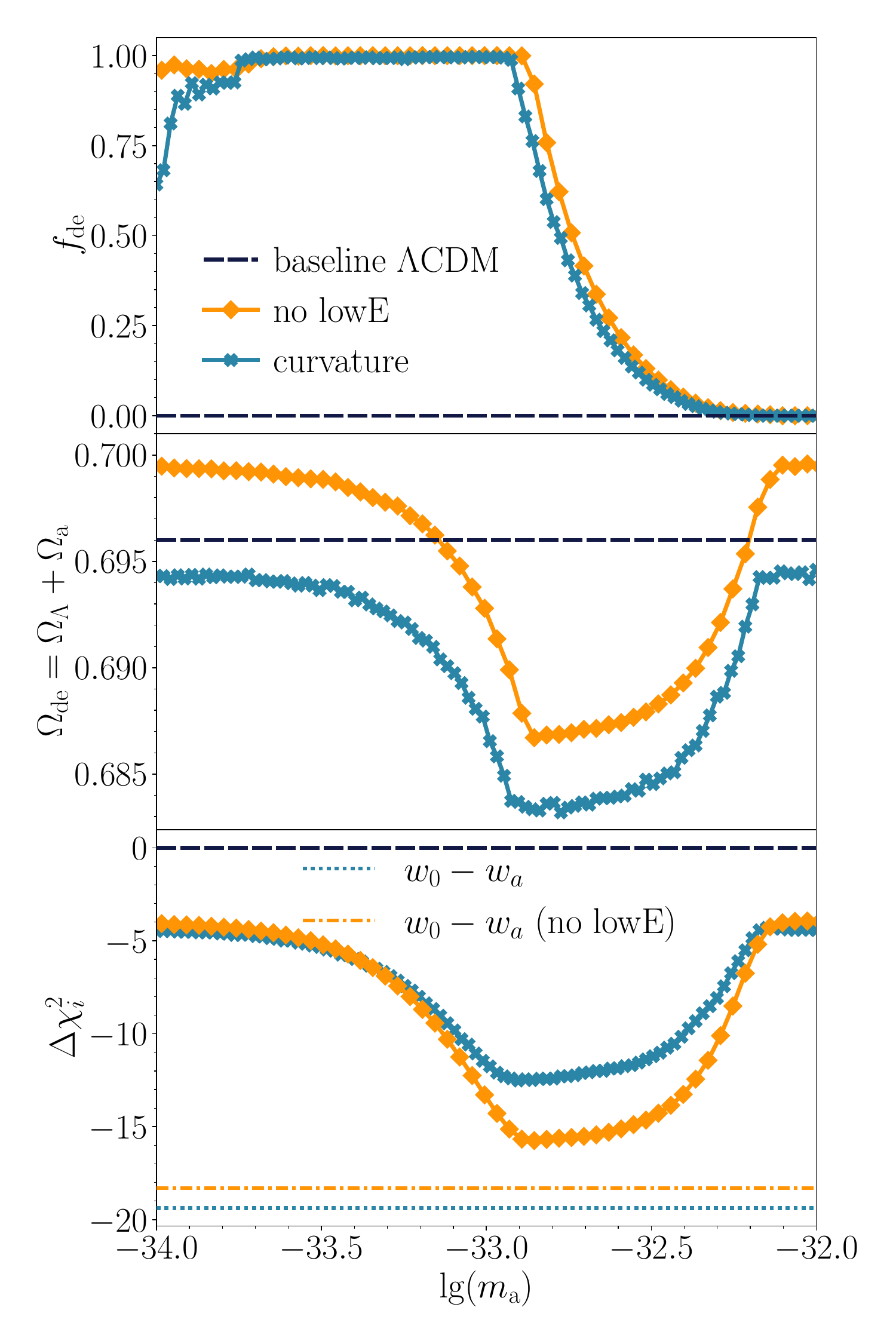}
    \caption{
    Likelihood profile as a function of axion mass or $\Delta\chi^2$ relative to the baseline $\Lambda$CDM model for the extended no lowE and curvature analyses (lower panel).  $\Delta\chi^2$ values are relative to the baseline $\Lambda$CDM  model but are given for $\Delta\chi^2_{\rm nolowE}$ in the former and the total for the latter.
    The two solutions prefer slightly different values of $\Omega_\de$ (middle panel) and the no lowE case allows larger axion dark energy fractions $f_\de$ than the curvature case (upper panel).   The no lowE case fits the rest of the SN+BAO+CMB data given $\Delta\chi^2_{\rm nolowE} \sim -16$ almost as well as the $w_0-w_a$  case (dot-dashed lines) whereas for the curvature case the total $\Delta\chi^2 \sim -12$ is notably worse (dotted lines) but still significant.
    }
\label{fig:profiles_extensions}
\end{figure}

In Fig.~\ref{fig:posterior_miragemass_resolution}, we show the posterior constraints in $f_\de, \DMeff/\rd,\rd$ when the lowE data is dropped.   Note that the SN+CMB constraints now widen to include more of the BAO favored regime reflecting the reduction in BAO+CMB tension.  This occurs mainly due to the increase in $\rd$ from the ability to lower $\Omega_c h^2$.
 
In this case the best fit at $\lgm =-32.9$ and $f_\de\approx 1$ has an improvement of $\Delta\chi^2_{\nolowlEE}=-15.8$, very close to the best fit  $w_\de(z)=w_0 + w_a z/(1+z)$ phantom model where $\Delta\chi^2_{\nolowlEE}=-18.3$ using CAMB\footnote{\href{http://camb.info}{http://camb.info}.  \textsc{CAMB} uses the Parameterized Post-Fried\-mann phenomenological approach \cite{Fang:2008sn} to avoid ghost and gradient instabilities which is not a microphysical model of phantom dark energy unlike our axion phantom mirage scenario.} for predictions (see Tab.~\ref{tab:dchi2_table} and Appendix \ref{sec:emulator_training}).  Thus  by allowing the optical depth to reionization to be raised to $\tau \approx 0.1$, the axion phantom mirage fits the rest of the CMB, SN and BAO almost as well.
This fit only degrades slightly across the phantom mirage mass range.  For example at $\lgm=-32.5$ and $f_\de \approx 0.12$
,  $\Delta\chi^2_{\nolowlEE}=-14.5$.  
Formally there is still only one extra parameter $f_\de$ over $\Lambda$CDM and $\Delta {\rm AIC}=-12.5$ compared with two for $w_0-w_a$, though the removal of lowE data should be considered a type of model complexity.
Notice that eliminating the source of BAO+CMB tension allows the best $f_\de$ at this mass to increase, which allows a better fit to the SN and BAO data due the axion phantom mechanism.  

These good fits to the SN, BAO and CMB lensing data sets are shown in Fig.~\ref{fig:SN_extensions}-\ref{fig:Clphiphi_phantomextensions} for the  $\lgm=-32.5$ case.  In fact the $\lgm=-32.5$ case has a  sharper upturn in SN magnitudes at the lowest redshifts than the global best fit $\lgm=-32.9$ case. It also allows a larger CMB lensing amplitude than even the fiducial model while still fitting the BAO with $\Omega_c h^2=0.1162$ 
vs.\ the fiducial value $0.12$ by raising $A_s$. 

A second type of solution to the BAO-CMB tension is to leave the sound horizon calibration fixed $\rd \approx \rd{}_{, \rm fid}$ but introduce spatial curvature $\Omega_K$ to change $\Dstar-\DMeff$ \cite{DESI:2025zgx,Chen:2025mlf}.  With just open $\Lambda$CDM, this has the effect of lowering $\Omega_m$ and raising $H_0 \rd$ through $H_0$ into better compatibility with BAO.
More generally curvature changes the relationship between 
$\DMeff$ and $\Dstar$ without altering $\Omega_c h^2$ or $\rd$.

With the addition of the axion phantom mirage, the same effect allows $\DMeff/\rd$ to become larger and bring it into compatibility with the BAO measurements
 as shown in Fig.~\ref{fig:posterior_miragemass_resolution}.  
 In Tab.~\ref{tab:dchi2_table}, we again give the best fit model at $\lgm=-32.9$ and $f_\de=0.85$ as well as at $\lgm=-32.5$ and $f_\de = 0.09$ where $\Omega_K \approx 0.003$.
 Because the addition of curvature does not change the $\rd$ inference from the CMB, it does not resolve the BAO+CMB tension quite as well as dropping lowE.  
 The improvement in $\Delta\chi^2$ reaches $-12.5$ and $-11.3$ whereas $\Delta {\rm AIC} = -8.5$ and $-7.3$ respectively for the two extra parameters.
We can see this in the comparison to BAO data in  Fig.~\ref{fig:BAO_extensions} where the high redshift end fits less well than the no lowE solution and in Fig.~\ref{fig:Clphiphi_phantomextensions} where it fails to raise the CMB lensing amplitude.

A third type of solution is to change the sound horizon, or correspondingly $\rd$, through a component of Early Dark Energy \cite{Poulin:2025nfb,SPT-3G:2025vyw}.  We leave this possibility to future work.

\section{Discussion}
\label{sec:discussion}

We have shown that the ability of axions with mass $m_\ax \gtrsim H_0$ to mimic a cosmological constant at high redshift and dark matter at low redshift causes a mirage of phantom dark energy and can explain the relative distances to SN and BAO.  In fact this effective dark energy can cross the phantom divide and provide an even sharper change with redshift if the true dark energy has an equation of state $w_\de>-1$ like quintessence, without causing any instabilities in the dark sector.

This resolution of the SN-BAO tension at $z \lesssim 1$ in $\Lambda$CDM fits essentially as well as calibrated thawing quintessence models.  Moreover it allows for a wide range of axion fractions from $\sim 5\%-100\%$ of the dark energy across this mass range.   

With axions alone, the BAO-CMB tension remains and is driven by the CMB determination of the cold dark matter density $\Omega_c h^2$ which largely determines the BAO distance to $z\sim 0.8$ and the CMB distance to recombination by calibrating the sound horizon or equivalently the BAO scale $\rd$.   In the truly phantom $w_0-w_a$ dark energy solutions, this is resolved by a stronger dark energy evolution at high $z$. On the other hand, this tension, being largely at high-$z$,  does not necessarily require a dark energy resolution.  

In particular, the CMB calibration of $\rd$ has become increasingly reliant on CMB lensing measurements but that requires knowledge of $\tau$, the optical depth to recombination, to fix $\Omega_c h^2$.   We show that the remaining tension between SN-BAO-CMB data can be resolved if $\tau \approx 0.1$.   
While this is not allowed by lowE polarization in the standard reionization history and power law inflationary power spectrum which imply $\tau= 0.0544 \pm 0.00755$ \cite{Planck:2018vyg}, more general models or systematic errors in the measurement could accommodate a higher $\tau$.    Indeed, JWST measurements of high-redshift galaxies 
\cite{2024ApJ...969L...2F,2022ApJ...938L..15C,2023arXiv230602465E,2023ApJS..265....5H} may  already be  hinting at earlier reionization (e.g.~\cite{Munoz:2024fas,2025Natur.639..897W}).
Ignoring the lowE polarization contribution to $\chi^2$, axions improve the rest of the fit to SN+BAO+CMB over $\Lambda$CDM  by $\Delta\chi^2 \sim -16$ and is comparable to the $w_0-w_a$ solution where $\Delta\chi^2 \sim -18$.   

Alternately, high redshift extensions to $\Lambda$CDM can also change the CMB inferences for $\rd$ and BAO distances. We show that a small curvature contribution $\Omega_K \sim 0.003$ can also relieve the BAO-CMB tension by adjusting the distance between $z\sim 0.8$ and recombination, bringing  the overall best fit to $\Delta\chi^2\sim -12.5$.    Modifying 
$\rd$ to solve the Hubble constant tension with the SHOES Cepheid-SN distance scale \cite{Riess:2025chq}
by introducing an ``early dark energy" scalar field that contributes near recombination can also relieve the the BAO-CMB tension.  This opens the intriguing possibility that a more general scalar sector could solve all of these tensions with $\Lambda$CDM simultaneously.  We leave these considerations to a future work.

\acknowledgments

  We thank Fei Ge for help with the CMB lensing data and Tom Crawford, Tanisha Jhaveri, Austin Joyce, and Tanvi Karwal for useful comments.
 R.L \& W.H. are supported by U.S.\ Dept.\ of Energy contract DE-SC0009924 and the Simons Foundation. 
 VM and YZ are supported by the Roman Project Infrastructure Team ``Maximizing Cosmological Science with the Roman High Latitude Imaging Survey" (NASA contracts 80NM0018D0004-80NM0024F0012). V.M. is also partially supported by the Roman Project Infrastructure Team ``A Roman Project Infrastructure Team to Support Cosmological Measurements with Type Ia Supernovae" (NASA contract 80NSSC24M0023). 
 Computing resources were provided by the  University of Chicago  Research
Computing Center through the Kavli Institute for Cosmological Physics and by Stony Brook Research Computing and Cyberinfrastructure, and the Institute for Advanced Computational Science at Stony Brook University through  the high-performance SeaWulf computing system.

\appendix

\section{Emulator Training Procedure}\label{sec:emulator_training}

\begin{table}[H]
    \centering
    \renewcommand{\arraystretch}{1.3}
    \begin{tabular}{lcc}\hline
        Parameter & Value \\\hline
        \texttt{accuracy\_boost} & 1.5 \\
        \texttt{l\_max\_scalar} & 7500\\
        \texttt{l\_accuracy\_boost} & 1.0\\
        \texttt{l\_sample\_boost} & 2.0\\
        \texttt{k\_eta\_max\_scalar} & 18000\\
        \texttt{do\_late\_rad\_truncation} & False\\
        \texttt{transfer\_kmax} & 10\\
        \texttt{transfer\_k\_per\_logint} & 130\\
        \texttt{transfer\_high\_precision} & True\\
        \texttt{accurateBB} & True

    \end{tabular}
    \caption{Settings for AxiECAMB} 
    \label{tab:AccuracySettings}
\end{table}

For the analyses in the main text we have both modified and emulated \textsc{AxiECAMB}.   The modification makes \textsc{AxiECAMB} solve the axion Klein-Gordon equation $\Box\phi= -m_\ax^2 \phi$ to the present for all axion masses in the quadratic potential approximation for axions. 

\begin{figure}[!ht]
    \centering
    \includegraphics[width=0.95\columnwidth]{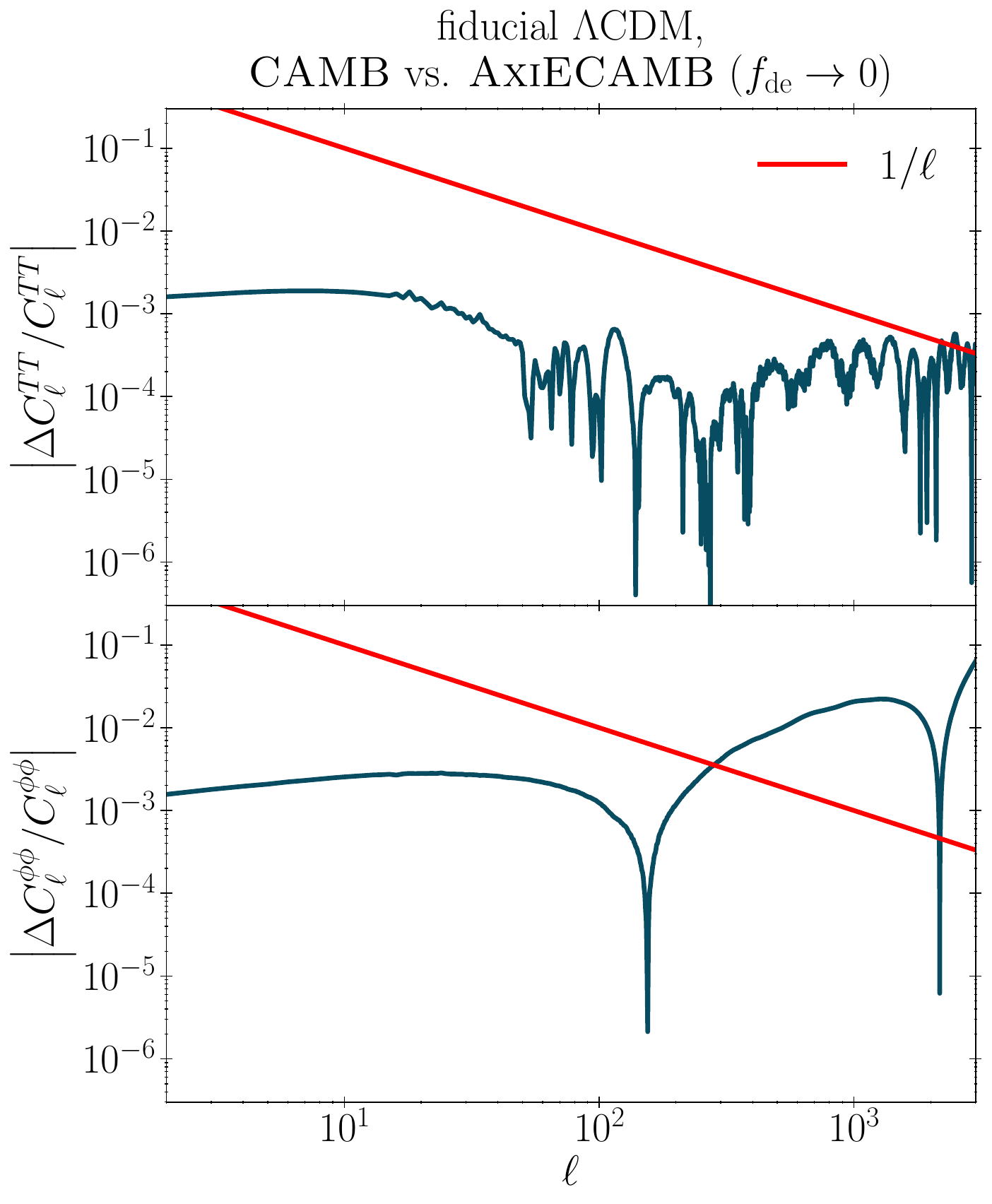}
    \caption{Comparison between AxiECAMB and CAMB  for $C_{\ell}^{TT}$ (top) and $C_{\ell}^{\phi\phi}$ (bottom) under the settings described in the main text. The relative difference for $TT$ is largely due to differing recombination codes, and for $\phi\phi$ to nonlinear power spectrum codes.  Deviations appear at high $\ell$ and the $1/\ell$ line approximates a cosmic variance limited measurement out to $\ell$.   The accuracy suffices for the datasets we use. }
    \label{fig:TTPPcomp}
\end{figure}

The accuracy settings for \textsc{AxiECAMB} are given in Tab.~\ref{tab:AccuracySettings}.
These are sufficient to make the accuracy of the $\chi^2$ for CMB likelihoods limited by the use of \textsc{RECFAST}
for recombination and takahashi \textsc{HALOFIT} \cite{Takahashi:2012em} for the nonlinear power spectrum in \textsc{AxiECAMB} as opposed to \textsc{COSMOREC} and mead2020 \textsc{HMCODE} \cite{Mead:2020vgs} for \textsc{CAMB v1.5.9} with accuracy settings specified in Ref.\ \cite{ACT:2025tim}.
For reference, the impact of these changes on the power spectra in the $\Lambda$CDM fiducial modal are shown in Fig.~\ref{fig:TTPPcomp} and mostly occur at very high $\ell$ compared with those of the measurements so that the errors there fall below the rough cosmic variance limits of $\Delta C_\ell/C_\ell \sim 1/\ell$ for the temperature anisotropy and  below the measurement noise for lensing (cf.~Fig.~\ref{fig:Clphiphi_examplemasses}).

There is a correspondingly small error in the computation of $\chi^2$ between \textsc{AxiECAMB} and \textsc{CAMB} that is well below $\Delta\chi^2=1$ as given in Tab.~\ref{tab:COSMORECMEADCOMP}.  By reverting \textsc{CAMB} to the older version of halofit and recombination used by \textsc{AxiECAMB} we see that the choice of recombination code mainly impacts the $\chi^2$ for Planck lite TTTEEE but only by  $\Delta\chi^2 \sim 0.1-0.2$ level. The choice of nonlinear power spectrum model affects mainly lensing but by $\Delta\chi^2 \lesssim 0.1$ if we run \textsc{CAMB} with mead2020 \textsc{HMcode} and COSMOREC for recombination.
  One notable addition is an extra $\Delta\chi^2\approx0.1$ in Planck lowT likelihood.  Fig.~\ref{fig:TTPPcomp} shows that the $C_\ell^{TT}$ predictions themselves are highly accurate so it should not affect parameter estimation.  In addition to the usual  linear scaling with $\Delta C_\ell^{TT}/C_\ell^{TT}$ for $\Delta \chi^2\ll 1$ rather than quadratically for $\Delta\chi^2\gg 1$ due to statistical fluctuations, this likely also reflects the enhanced changes in $\chi^2$ for a given accuracy of computation due to $\Lambda$CDM being a somewhat poor fit to the lowT data. 
While these together set a floor in the accuracy of our computations it is sufficient for the datasets we consider.  It also sets the level of diminishing returns that motivate the precision settings
in Tab.~\ref{tab:AccuracySettings} and motivate the accuracy goals of the emulator as we shall see next.

\begin{table}[H]
    \centering
    \renewcommand{\arraystretch}{1.3}

{
    \begin{tabular}{ccccccc}\hline
  Nonlinear & Recombination & \pbox{10cm}{TTTEEE} & LowT  & LowE & $\phi\phi$  \\\hline
 mead2020 & COSMOREC & 0.31 & -0.11 & 0.00 & 0.06\\
 mead2020 & RECFAST & 0.15 & -0.11 & 0.01 & 0.11\\
 takahashi & COSMOREC & 0.31 & -0.11 & 0.00 & 0.07\\
 takahashi & RECFAST  & 0.15 & -0.11 & 0.01 & 0.12\\\hline
    \end{tabular}
 
    }
    \caption{$\Delta\chi^2$ changes for  CAMB v1.59 relative to AxiECAMB on the fiducial $\Lambda$CDM model (top row, ACT accuracy settings).
    For Planck lite TTTEEE, the difference between COSMOREC and RECFAST contributes more than half of $\Delta\chi^2$ as shown by switching the two in CAMB. 
    For LowE, the differences are negligible. For LowT, $\Delta\chi^2\approx -0.1$ between  CAMB and AxiECAMB across all the test cases.  Reverting mead2020 to takahashi in CAMB changes the lensing likelihood by less than 0.1 and others entirely negligibly.}
    \label{tab:COSMORECMEADCOMP} 
\end{table}

We then emulate the \textsc{AxiECAMB} calculations following  the architectures and training strategies developed in Ref.~\cite{Zhong:2024xuk,Saraivanov:2024soy,Zhu:2025jim}. To perform a thorough analysis of the axion cases with SN, BAO and CMB datasets, the required data vectors to be emulated include the CMB primary power spectra $C_{\ell}^{XY}$ ($X,Y\in T,E$), the lensing power spectrum, $C_{\ell}^{\phi\phi}$,  $H(z)$, and the BAO drag scale $\rd$.
We also emulate the mapping of $\theta_*$ to $H_0$ in order to sample the posterior in $\theta_*$, which is directly constrained by the CMB.

We adopt the Residual MLP (ResMLP) with PCA architecture for $H(z)$ and $C_{\ell}^{\phi\phi}$. For the mapping from $\theta_*$ to $H_0$, we train a ResMLP with smaller dimensions and size compared to the one above. For CMB primary power spectra, we apply a combination of ResMLP and Convolutional Neural Network (CNN). A Gaussian Processing (GP) neural network is found to be sufficient for $\rd$, as we can simply train this using  $\Lambda$CDM parameters given that axions with the mass range in our work have negligible energy density at recombination. Transverse comoving distances are computed by performing the integration in $H(z)$; in general
\begin{equation}
    D_M(z)=\frac{1}{\Omega_K H_0^2} \sinh\left[ {\Omega_K H_0^2 \int_0^z \frac{ dz}{H(z')}}\right],
    \label{eqn:DMz}
\end{equation}
with the flat limit of $\Omega_K\rightarrow 0$ of $D_M=\int dz/H(z)$.  We adopt a composite-Simpson integration algorithm to obtain results from the Hubble parameter computed from the emulator.

The BAO radial distance $D_H(z)=1/H(z)$ and volume weighted $D_V(z)=[z D_M^2(z) D_H(z)]^{1/3}$ follow directly.   Supernova magnitudes $m$, corrected for peculiar velocities, are compared to models through the distance modulus
\begin{equation}
\mu = m-M =5 \log_{10}\left[\frac{ (1+z)D_M(z)}{{1 {\rm Mpc}}}\right]+25,
\end{equation}
with the absolute magnitude $M$ marginalized in the likelihood or maximized in plots.  

\begin{figure}[!ht]
    \centering
\includegraphics[width=1\columnwidth]{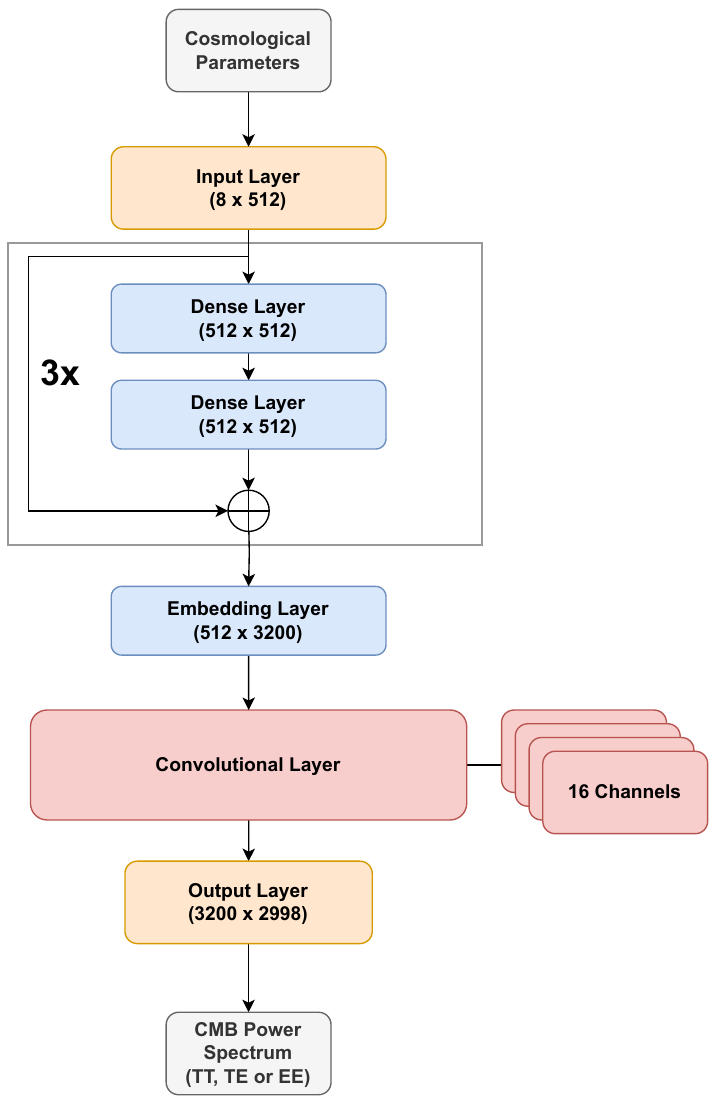}
    \caption{Architecture of the 1D convolutional neural network.}
    \label{fig:CNN}
\end{figure}

A complete list of emulator configurations used are summarized in Table.~\ref{tab:Architectures}. The CNN architecture is shown in Fig.~\ref{fig:CNN}, with the channel number of the CNN fixed at $16$.
\begin{table}[H]
    \centering
    \renewcommand{\arraystretch}{1.3}
    
    \begin{tabular}{lcc}\hline
        Observable & Architecture \\\hline
        $C_{\ell}^{XY}$, $X,Y\in T,E$ & ResMLP+CNN \\
        $C_{\ell}^{\phi\phi}$ & ResMLP+PCA\\
        $H(z)$ & ResMLP+PCA\\
        $\rd$ & Gaussian Processing\\
        $\theta_*\rightarrow H_0$ & ResMLP\\
         Distance Observables & Derived from $H(z)$
    \end{tabular}
    
    \caption{Architectures for each observable.}
        \label{tab:Architectures}
\end{table}

To cover a sufficiently large parameter space for a thorough analysis, we adopt uniform sampling among the priors in Table.~\ref{tab:prior}. The training prior is slightly larger than the testing prior to avoid edge effects.
\begin{table}[H]
    \centering
    \renewcommand{\arraystretch}{1.1}
    {
    \begin{tabular}{lcc}\hline
        & Uniform & Uniform\\
        Parameter & Train & Test\\\hline
        $\Omega_b h^2$  & $[0.002,0.04]$ & $[0.006,0.038]$ \\
        $\Omega_c h^2$   & $[0.03,0.24]$  & $[0.04,0.23]$  \\
        $H_0$              & $[55,85]$      & $[60,80]$   \\
        $\tau$           & $[0.005,0.105]$  & $[0.01,0.15]$  \\
        $\ln(10^{10}A_s)$ & $[1.61,3.6]$    & $[1.7,3.5]$    \\
        $n_s$              & $[0.7,1.3]$     & $[0.8,1.2]$  \\
        $\lgm$ & $[-35,-31]$  & $[-34,-31.5]$  \\
        $\Omega_\ax h^2[\lgm \leq -32]$ & $[0,0.4]$  & $[10^{-8},0.38]$     \\
        $\Omega_\ax h^2[\lgm > -32]$ & $[0,0.21]$  & $[10^{-8},0.19]$     \\
        \hline
    \end{tabular}
    }
    \caption{Parameter ranges for the training and testing parameter sets for the uniform and Gaussian sampling methods.  }
    \label{tab:prior}
\end{table}
For the models with curvature, we further extend the space to analyze  non-zero $\Omega_K\in[0,0.004]$, with trimmed priors on $\tau\in[0.04, 0.12]$, $ \ln(10^{10}A_s)\in[3, 3.3]$, $n_s\in[0.92, 0.99]$ and 
$\lgm\in[-34, -32]$, which includes the region of interest for fitting the datasets.  

To well emulate the region where $\fa$ is small, when generating training datasets, we oversample in the low $\Omega_{\ax}h^2$ region by applying an additional uniform sampling in logarithmic measure of $\log_{10}{(\Omega_{\ax}h^2)}\in[-5,-1]$. We also oversample the region where $\lgm>-32.5$ since the relationship between cosmological parameters and data vectors is more non-linear in this region due to more oscillations in the background evolution of the axion field. Furthermore, we sample directly at $\Omega_{\ax}h^2=10^{-8}\approx0$ to force the emulator to learn the $\Lambda$CDM limit.

\begin{figure}[!ht]
    \centering
    \includegraphics[width=0.95\columnwidth]{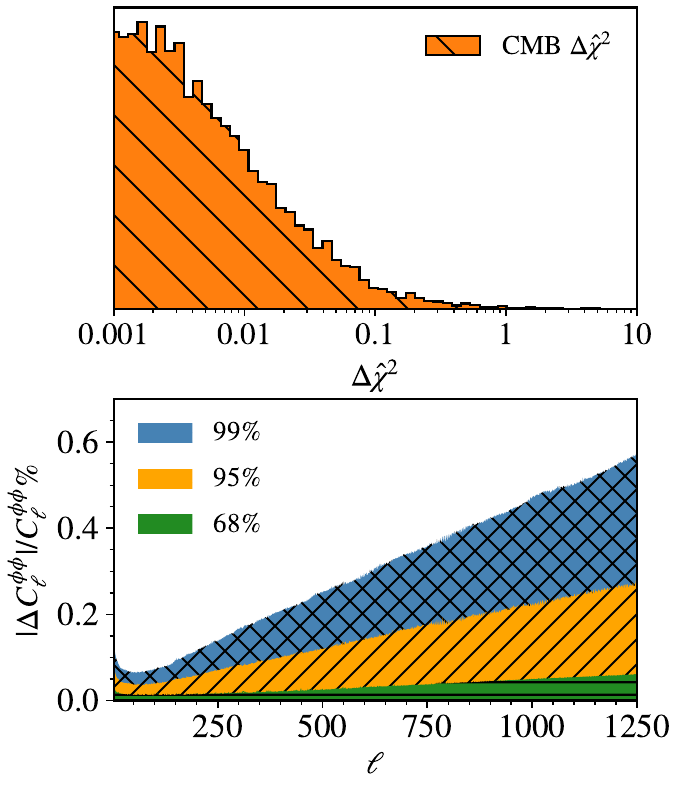}
    \caption{Top: Distribution of $\Delta{\hat \chi}^2$ of the CMB primaries emulators upon testing on Planck TTTEEE lite likelihood. Only $1\%$ points are outliers with $\Delta{\hat\chi^2}>0.2$. Bottom: Distribution of percent difference between emulator and AxiECAMB outputs of  $C_{\ell}^{\phi\phi}$. Within the range $\ell\leq1250$, which is the ACT lensing extended multipole range, whereas we use the baseline ACT range $\ell \leq 763$ in our analysis, we can see that the percent error of the emulation is suppressed to below $1\%$ across the range of ACT lensing likelihood.}
    \label{fig:CMBloss}
\end{figure}

For $\rd$ and $\theta_*\rightarrow H_0$ mapping emulators, we train with the simple Mean Square Error (MSE) loss function of the output directly. For $H(z)$ and $C_{\ell}^{\phi\phi}$ emulators, we train the models under MSE of the PCA transformed outputs. Since those data vectors are either short in size or structurally simple, a crude MSE loss function and simple training strategies are sufficient.

The more challenging training is for CMB primaries $C_{\ell}^{XY}$ as they have more complicated structures. To train them to be as precise as possible compared to the direct output from AxiECAMB, we set the training loss function, separately for each $C_{\ell}^{XY}$, to be 
\begin{equation}
    {\rm loss} = \Big\langle\Big(\sum_{\ell}\Delta \Tilde{C}_{\ell}\Tilde{C}^{-1}\Delta \Tilde{C}_{\ell}\Big)^{1/2}\Big\rangle \, .
\end{equation}
$\Delta\Tilde{C}_{\ell}$ is the difference in the rescaled quantity as $\Tilde{C}_{\ell}= C_{\ell} e^{2\tau}/A_s$, 
where the fiducial model picked at near the Planck 2018 best fit.
The covariance matrix $\tilde C$ is set to be the corresponding diagonal part of a cosmic-variance covariance matrix for rescaled quantities with $\ell\leq3000$ for the sake of training. 
This rescaling technique is tested in Ref.~\cite{Zhu:2025jim}. The increasingly tighter constraints as $\ell$ increases under the cosmic-variance assumption forces the training to learn the detailed structure at high $\ell$.

For testing, we compute the quantity $\Delta\hat{\chi}^2 = \sum \Delta C_{\ell}^{ XY}C^{-1}_{XY,X'Y'}\Delta C_{\ell}^{X'Y'}$, with $X,Y,X',Y'\in T, E$, using the Planck lite binning and covariance matrix files. We then evaluate the median and distribution of this quantity across a testing set. Note that $\Delta\hat{\chi}^2$ does not involve the Planck data itself and therefore does not get enhanced if the model is a bad fit to the data or from the noise fluctuations of a good fit.

Upon testing, our emulators have almost consistently below $0.5\%$ error in $D_M(z)$ and $H(z)$ across the redshift range $z\in[0,3]$. The $\rd$ GP emulator has around $0.02\%$ error, and the emulator for mapping from $\theta_*$ to $H_0$ around $0.04\%$ error. Both of these derived parameter emulators have errors one order of magnitude lower than the $1\sigma$ deviation from the Planck measurements. For the CMB primary power spectra, our emulators are trained to have the fraction of outliers with $\Delta\hat{\chi}^2>0.2$ to be around $1\%$ under Planck lite TTTEEE likelihood within our testing range, shown in Fig.~\ref{fig:CMBloss}. The CMB lensing power spectrum emulator is trained to have $ < 1\%$ error across the $\ell$ range of ACT lensing. The training for the baseline spatially flat cases uses $2.5$ million training points. We then develop another emulator for the curvature analysis with an addition of $0.8$ million points sampled in the range described above, and the models reach a similar level of precision.

\vfill
\bibliographystyle{apsrev4-1}
\bibliography{phantom}

\end{document}